# SPEAKER IDENTIFICATION FROM YOUTUBE OBTAINED DATA

Nitesh Kumar Chaudhary[1] and Shraddha Srivastav[2]

[1]Department of Electronics & Communication Engineering, LNMIIT, Jaipur, India
[2]Bharti School Of Telecommunication, IIT, Delhi, India

## ABSTRACT

*An efficient, and intuitive algorithm is presented for the identification of speakers from a long dataset (like YouTube long discussion, Cocktail party recorded audio or video).The goal of automatic speaker identification is to identify the number of different speakers and prepare a model for that speaker by extraction, characterization and speaker-specific information contained in the speech signal. It has many diverse application specially in the field of Surveillance , Immigrations at Airport , cyber security , transcription in multi-source of similar sound source, where it is difficult to assign transcription arbitrary. The most commonly speech parameterization used in speaker verification, K-mean, cepstral analysis, is detailed. Gaussian mixture modeling, which is the speaker modeling technique is then explained. Gaussian mixture models (GMM), perhaps the most robust machine learning algorithm has been introduced to examine and judge carefully speaker identification in text independent. The application or employment of Gaussian mixture models for monitoring & Analysing speaker identity is encouraged by the familiarity, awareness, or understanding gained through experience that Gaussian spectrum depict the characteristics of speaker's spectral conformational pattern and remarkable ability of GMM to construct capricious densities after that we illustrate 'Expectation maximization' an iterative algorithm which takes some arbitrary value in initial estimation and carry on the iterative process until the convergence of value is observed We have tried to obtained 85 ~ 95% of accuracy using speaker modeling of vector quantization and Gaussian Mixture model ,so by doing various number of experiments we are able to obtain 79 ~ 82% of identification rate using Vector quantization and 85 ~ 92.6% of identification rate using GMM modeling by Expectation maximization parameter estimation depending on variation of parameter.*

## KEYWORDS

*MFCC, Vector Quantization (VQ), Gaussian Mixture Model (GMM), K-mean, Maximum Likelihood, Expectation maximization.*

## 1. INTRODUCTION

Speaker identification have two categories: text-dependent and text-independent. Essentially, we are more interested and involved in the research of text independent speaker identification / verification with the reason that it doesn't impose as a necessity and demands regarding the utterances obtained from speaker that means there is no restriction over the words , it can be anything in any order, so the first basic steps involve the feature extraction from speech sample in the enrolment process of a speaker , extracted feature are collected in the database as a training data utterances. For the better accuracy, every time we are updating our training dataset while preparing training dataset for new utterances with previous stored data in our database, in this way model of each speaker is trained. Now in identification process with the help of probabilistic model a measurement of identification has to be done, the feature vectors of testing utterances is measured with feature vectors of training dataset and decision has to be made whether it belongs to a group of dataset or not.

## 2. FEATURE EXTRACTION FROM VOCAL TRACT

Human's vocal tract, the airway used in the production of speech is the organs above the vocal folds, especially the passage above the larynx, including the pharynx, mouth, and nasal cavities. which is formed of the oral part (pharynx, tongue, lips, and jaw) , olfactory nerves, and the nasal tract. When the glottal pulses signal generated by the vibration of the vocal folds passes through the vocal tract, it is modified. Human's vocal tract is performing like a filter, and its frequency characteristics is dependent upon the resonance peak from the vocal tract and vocal tract configuration can be obtained from the spectral shape such as formant position and spectral inclination of the speech signal. These features can be obtained from the spectrogram of the speech signal and we are using Mel-Frequency Cepstral Coefficients (MFCC) features in speaker identification as it combines the advantages of the cepstrum analysis with a perceptual frequency scale based on critical bands. Although the speech signal is non-stationary, but can be assumed as stationary for a short duration of time, so analysis is done by framing the speech signal; the frame width is about 20–30 milliseconds, and the frames are shifted by about 10 milliseconds.

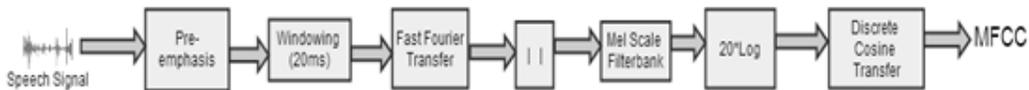

Diagram 1. MFCC algorithm for feature extraction

The number of feature vector that we get from utterances is usually large so for computation and the number of feature vectors can diminished without lose by K-means clustering method. This results in a small set of vectors called as codebook vector and a codebook is obtained for each speaker utterances using K-means. A codebook consist of many different vectors, which depict the important characteristics of each speaker. Each codebook is obtained as follows: Given a set of training set of MFCC feature vectors of 16-point vector for each frame of the utterance, which represent the speaker, find a decomposition of the feature vector space. Each decoposed region contains a cluster of vectors, which depict the same kind of basic sound. This region is represented by the centroid vector, which is the vector, which causes the minimum distortion when vectors in the region are mapped to it. Thus, each speaker has a codebook with a number of centroids which is prepared with the help of K-mean Clustering.

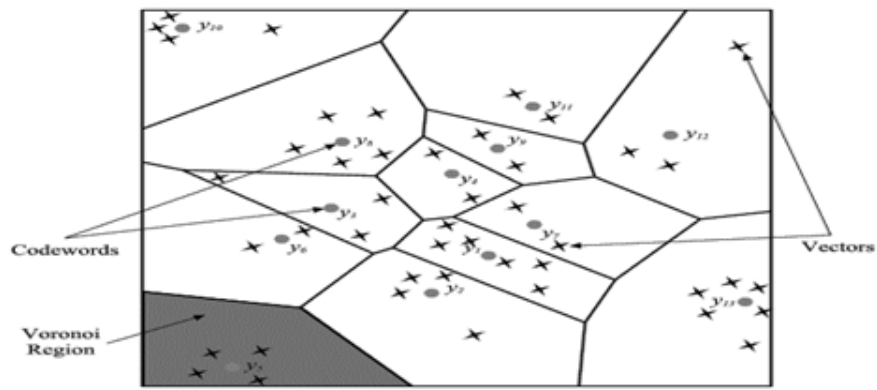

Diagram 2. Codebook generation using K-mean Clustering

## 3. MODELING USING VECTOR QUANTIZATION AND IDENTIFICATION

Vector quantization (VQ) is one of the simplest text-independent speaker models. VQ is often used for computational speed-up techniques and lightweight practical implementations. Vector quantization (VQ) is a ancient well-known quantization skills from signal processing which allows the modeling of probability density functions by the classification of prototype vectors. VQ is generally used by classifying a large se of (MFCC)feature vector dataset in small groups vectors having same number of points closet to the denser value i.e. means of codebook of training dataset

For Identification average quantization distortion is computed for the test utterance feature vectors by X = {x1,x2....xT}and the reference vectors by R={r1,r2….rk} .

$$D_Q(X, \mathscr{R}) = \frac{1}{T} \sum_{t=1}^{T} \min_{1 \leq k \leq K} d(x_t, r_k)$$

Where $d(.,.)$ is a distance measure such as the Euclidean distance $\|x_t-r_k\|$. A smaller value $D_Q$ indicates higher likelihood for X and R originating from the same speaker.

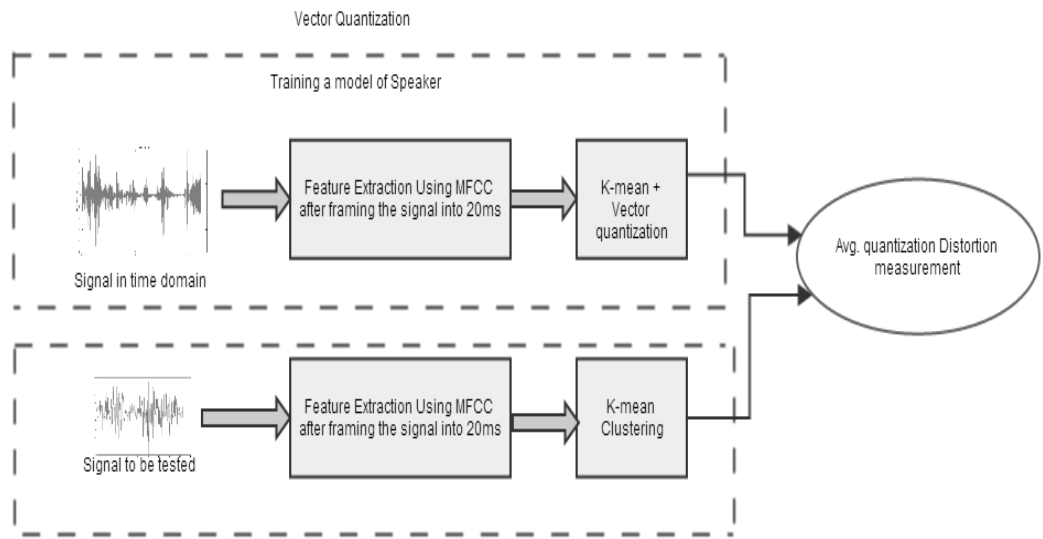

Diagram 3

Above Diagram1 shows speaker identification using Vector quantization modeling, the upper region is enrolment process while in lower region feature has been extracted from test utterance.
Diagram2 shows the Vector quantization distortion measurements plot when the speaker identification's test has been performed for 8 speakers. For K = 16 variations of distortion measurement from speaker i to speaker i utterances lies between (3.4561 to 7.0723) and from speaker i to speaker j lies between (13.1892 to 29.9038), where i, j $\epsilon$ (1 to 8).

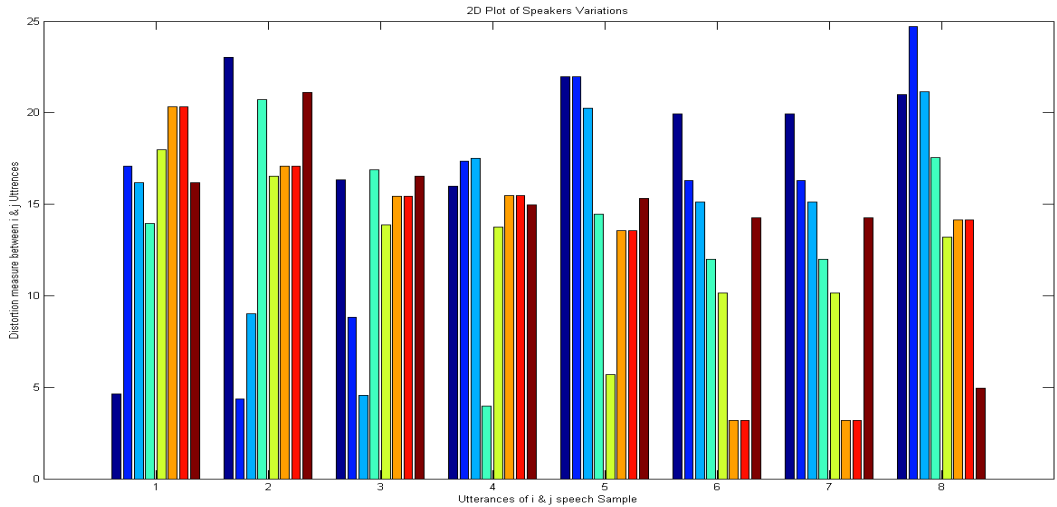

Diagram 4

**Table1**. Performance of Vector quantization modeling with MFCC feature extraction

| Training Utterances | No. of Clustering (K mean) | Identification rate (%) |
|---|---|---|
| 8 Speaker utterances | K = 32 | 82 |
| 8 Speaker utterances | K = 16 | 82 |
| 8 Speaker utterances | K = 8 | 80 |

Total Number of speaker's utterances = 75

## 4. MODELING USING GAUSSIAN MIXTURE MODEL

A Gaussian Mixture Model (GMM) is a parametric probability density function represented as a weighted sum of Gaussian component densities. It is generally used as a parametric model of the probability distribution of continuous spectrum. GMM parameters are computed by 'Expectation maximization' an iterative algorithm which takes some arbitrary value in initial estimation and carry on the iterative process until the convergence of value is observed and the whole Gaussian mixture model is defined by these parameters namely mean vectors $(\mu_i)$, *covariance matrices*$(\sigma_i)$ and *mixture weights*$(p_i)$ from all different components, and the weighted sum of M Component density is given by

$$P(x/\lambda) = \sum_{i=1}^{M} p_i \cdot g(x/\mu_i, \sigma_i)$$

where $\chi$ is a D-dimensional continuous-valued feature vector (in our case 16 dimensional), $p_i$, i = 1, . . . ,M, are the mixture weights, and $g(\chi/\mu_i,\sigma_i)$ are component densities. Each component density is a D-dimensional Gaussian function of the form,

$$g(x/\mu_i, \sigma_i) = \frac{1}{\sqrt{2\pi^D \sigma_i}} \exp\left\{ -\frac{1}{2}(x-\mu_i)' \sigma_i^{-1}(x-\mu_i) \right\}$$

GMM based speaker identification model is shown below, the upper region is extraction of features from training dataset and preparing a mixture model using expectation maximization while in lower region feature has been extracted from test utterance, and finally with GMM parameter and extracted features from test utterances identification measurement has been done by using probability generation function

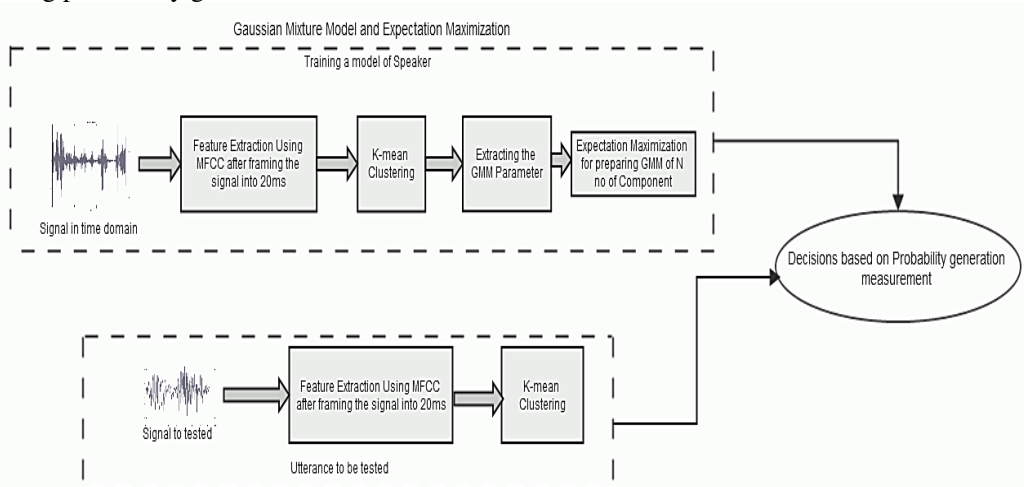

Diagram 5

One of the powerful attributes of the GMM is its ability to form smooth approximations to arbitrarily shaped densities. There are several techniques available for estimating the parameters of a GMM i.e ($\mu$, $\sigma$, $p$) and most popular and well-established method is maximum likelihood (ML) estimation. Since the equation of joint probability is nonlinear function of parameter $\lambda$ so we can't solve it easily because number of unknown variables are more than number of equations , so the parameter of ML is estimated iteratively by *expectation maximization* . The basic idea of the EM algorithm is, beginning with an initial model $\lambda_i$ to estimate a new model $\lambda_j$ such that $p(\chi/\lambda_j) \geq p(\chi/\lambda_i)$, where $j > i$ . The new model then becomes the initial model for the next iteration and the process is repeated until some convergence threshold is reached, so by using number of utterances = 10 can give you a good approximated parameters ($\mu$, $\sigma$, $p$) of GMM.

**Table2.** Performance of GMM with MFCC feature based identification

| Training data in seconds | Testing data in seconds | No. of component | No. of Iterations for EM | Rate of correct identification (%) |
|---|---|---|---|---|
| 6 | 6 | 2 | 6 | 78.8342 |
| 6 | 6 | 4 | 6 | 76.6654 |
| 12 | 6 | 4 | 8 | 83.6723 |
| 20 | 6 | 4 | 10 | 84.4348 |
| 30 | 4 to 10 | 4 | 12 | 87.4382 |
| 30 | 4 to 10 | 5 | 12 | 89.5630 |
| 60 | 4 to 10 | 6 | 12 | 92.6643 |
| 120 | 4 to 10 | 7 | 14 | 92.1273 |

Total Number of speaker's utterances = 75

## 5. CONCLUSION

In this paper, we have introduced a text-dependent speaker identification system, we have investigated that Gaussian mixture models (GMMs) have proven extremely successful for text-independent speaker identification for long dataset of different speakers. Identification performance of the Gaussian mixture speaker model is insensitive to the method of model initialization however we have estimated the parameter using expectation maximization and Identification rate is very sensitive to number of cluster and number of iteration of expectation maximization we have performed the experiments in Matlab R2014a (*The Language of Technical Computing*), this model is currently being evaluated on a 75 speaker's utterances and the results indicate that Gaussian mixture models provide a robust speaker representation for the difficult task of speaker identification using corrupted, unconstrained speech of cocktail party or YouTube dataset The models are computationally inexpensive and easily implemented on a real-time platform.

**REFERENCES**

[1] T.Matsui and S. Furui, "Likelihood normalization for speaker verification using a phoneme- and speaker-independent model," Speech Communication, vol. 17, no. 1-2, pp. 109–116, 1995.

[2] D. A. Reynolds, A Gaussian mixture modeling approach to text independent speaker identification, Ph.D. thesis, Georgia Institute of Technology, Atlanta, Ga, USA, September 1992.

[3] D. A. Reynolds, "Speaker identification and verification using Gaussian mixture speaker models," Speech Communication, vol. 17, no. 1-2, pp. 91–108, 1995.

[4] J. Attili, M. Savic, and J. Campbell. "A TMS32020-Based Real Time, Text-Independent, Automatic Speaker Verification System," In International Conference on Acoustics, Speech, and Signal Processing in New York, IEEE, pp. 599-602, 1988

[5] Levinson, S . E., L. R. Rabiner, A. E. Rosenberg and J. G. Wilson, "Interactive Clustering Techniques for Selecting Speaker-Independent Techniques for Selecting Speaker-Independent Reference Templates for Isolated Word Recognition," IEEE Trans. ASSP. Vol. 27, pp. 134-141,1979.

[6] R. Mammone, X. Zhang, R. Ramachandran, "Robust Speaker Recognition", IEEE Signal Processing Magazine, September 1996.

[7] F. Soong, E. Rosenberg, B. Juang, and L. Rabiner. "A Vector Quantization Approach

[8] Campbell, J., 1997. Speaker recognition: a tutorial. Proc. IEEE 85 (9),1437–1462.

[9] Hatch, A., Stolcke, A., Peskin, B., 2005. Combining feature sets with support vector machines: application to speaker recognition. In: The 2005 IEEE Workshop on Automatic Speech Recognition and Understanding (ASRU), November 2005, pp. 75–79.

[10] Reynolds, D. "Speaker Verification Using Adapted Gaussian Mixture Models." Digital Signal Processing 10.1-3 (2000): 19-41. Print.

[11] K. Chen and L. Wang and H. Chi, \Methods of combining multiple classifiers with different features and their application to text-independent speaker identification," International Journal of Pattern Recognition and Artificial Intelligence, Vol. 11, no. 3, pp. 417{445, 1997.

[12] Linde, Y., A. Buzo, and R. Gray (1980). An algorithm for vector quantizer design. IEEE Transactions on Communications, 28(1), 84–95.

[13] Lu, X. and J. Dang (2008). An investigation of dependencies between frequency components and speaker characteristics for text-independent speaker identification. Speech Communication, 50(4), 312–322.

[14] Madikeri, S. R. and H. A. Murthy (2011a). Mel filter bank energy-based slope feature and its application to speaker recognition. In Proc. National Conference on Communications, 1–4.

[15] NIST (2004). The NIST year 2004 speaker recognition evaluation plan. http://www.itl.nist.gov/iad/mig/tests/sre/2004/index.html.

[16] Sha, F. and L. K. Saul, Large margin gaussian mixture modeling for phonetic classification and recognition. volume 1. IEEE, 2006.

**AUTHORS**

Nitesh Kumar Chaudhary Final Year undergraduate (Electronics & Communication) student at LNMIIT, jaipur (India). My Research interest includes audio Signal Processing, Speaker Authentication & Identification , Machine Learning (Probabilistic model). I have worked as a Visiting Summer Researcher at Newyork University , Summer Research Fellow at Indian Institute of Science, Bangalore and worked as Research Intern at Indian Institute Of Technology ,Delhi a world class program organised by Bharti School Of Telecommunication , Foundation of Innovation and Technology Transfer.

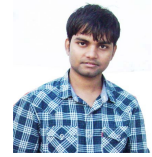

More Info : https://sites.google.com/site/nkcniteshkumarchaudhary/

Shraddha Srivastav Post Doc. Researcher at Bharti School of Telecommunication, Indian Institute of Technology, Delhi, PhD from University College Cork, Ireland (UCC)